\documentclass{article}

\usepackage{authblk}
\usepackage{graphicx}
\usepackage{hyperref}

\usepackage{caption}
\usepackage{subcaption}

\author[1]{Denis Kutnar}
\author[1]{Bas H.M. van der Velden}
\author[1]{Marta Girones Sanguesa}
\author[2]{Mirjam I. Geerlings}
\author[3]{J. Matthijs Biesbroek}
\author[1]{Hugo J. Kuijf}

\affil[1]{Image Sciences Institute, UMC Utrecht}
\affil[2]{Julius Center for Health Sciences and Primary Care, UMC Utrecht}
\affil[3]{Department of Neurology and Neurosurgery, UMC Utrecht}

\title{MixLacune: Segmentation of lacunes of presumed vascular origin}
\date{5 August 2021}

\begin{document}

\maketitle

\section{Introduction}
Lacunes of presumed vascular origin are fluid-filled cavities of between 3 – 15~mm in diameter \cite{Wardlaw2013,Pantoni2010}. They are visualized as a hypointense cavity on Fluid-attenuated inversion recovery (FLAIR) and T1-weighted imaging, usually with a hyperintense rim on FLAIR imaging \cite{Wardlaw2013,DeGuio2016}. Quantification of lacunes relies on manual annotation or semi-automatic / interactive approaches; and almost no automatic methods exist for this task \cite{DeGuio2016,Zhao2021}. Initial work by Ghafoorian \emph{et al.} \cite{Ghafoorian2017} presented a method for detection of lacunes with a deep multi-scale location-aware 3D convolutional neural network (CNN). Preliminary results by Ooms \cite{Ooms2021} suggest that segmentation of lacunes is feasible with a U-Net \cite{Ronneberger2015} CNN.

In this work, we present a two-stage approach to segment lacunes of presumed vascular origin: (1) detection with Mask R-CNN \cite{He2017} followed by (2) segmentation with a U-Net CNN.

\section{Material and methods}
\subsection{Data}
Data originates from Task 3 of the ``Where is VALDO?'' challenge (\url{https://valdo.grand-challenge.org/}) and consists of 40 training subjects. For each subject, a T1-weighted, T2-weighted, and FLAIR image are provided; including manual annotations of lacunes made by two different observers.

\subsubsection{Lacune prevalence map}
Lacunes do not occur randomly throughout the brain, but have preferential locations. They occur mostly in the basal ganglia and the white matter of the internal capsule and pons; and rarely at other locations \cite{Marti-Vilalta2004}. To exploit this prior knowledge, a lacune prevalence map was created that shows the expected distribution of lacunes throughout the brain; to be used as a mask on the final results.

\begin{figure}[tbh]
\centering
\includegraphics[width=\textwidth]{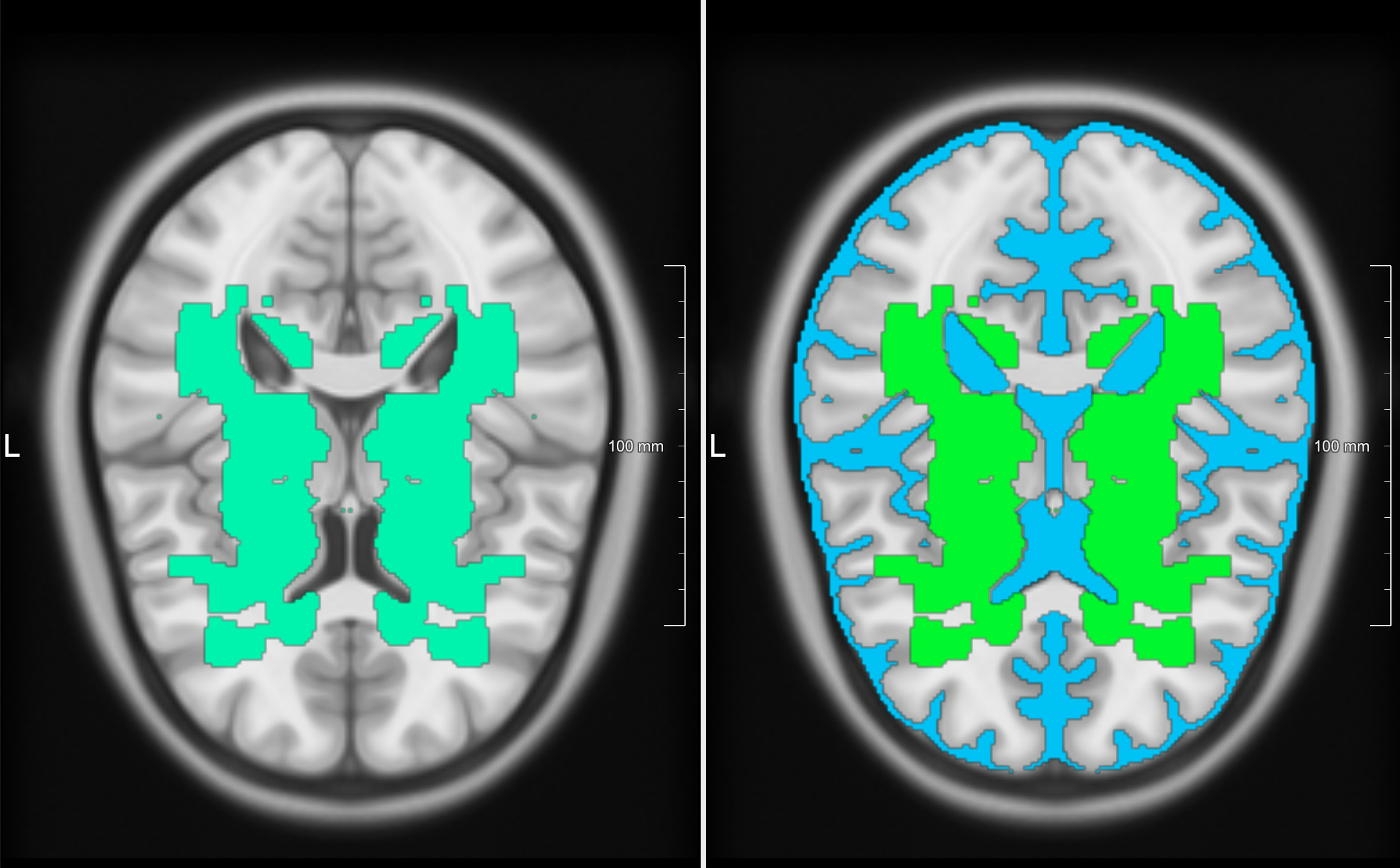}
\caption{The final lacune prevalence map in MNI152-space is shown on the left. Ventricles and extracerebral CSF were removed from the map with a CSF mask (light-blue on the right).}
\label{fig:fig1}
\end{figure}

From the SMART-MR study \cite{Biesbroek2013}, we included 98 subjects for which manual segmentation of lacunes were available; and combined this with the 40 training subjects of the VALDO challenge. All images were registered to the MNI-152 \cite{Fonov2011} atlas to create a lacune prevalence map. The resulting map (Figure \ref{fig:fig1}) was made symmetric by mirroring, dilated by 7~mm in all directions, and the ventricles and CSF were removed.

\subsection{Method}
Lacunes are hard to find, because of their small size, versatile shape, and sparse occurrence in the brain \cite{Wardlaw2013,DeGuio2016,Zhao2021}. For this reason, a two-stage segmentation approach was chosen. In the first step, Mask R-CNN \cite{He2017} localizes possible lacunes. In the second step, a U-Net \cite{Ronneberger2015} segments lacunes based on the locations suggested by the Mask RCNN. 

\subsubsection{Full prediction pipeline}
\begin{figure}[tbh]
\centering
\includegraphics[width=\textwidth]{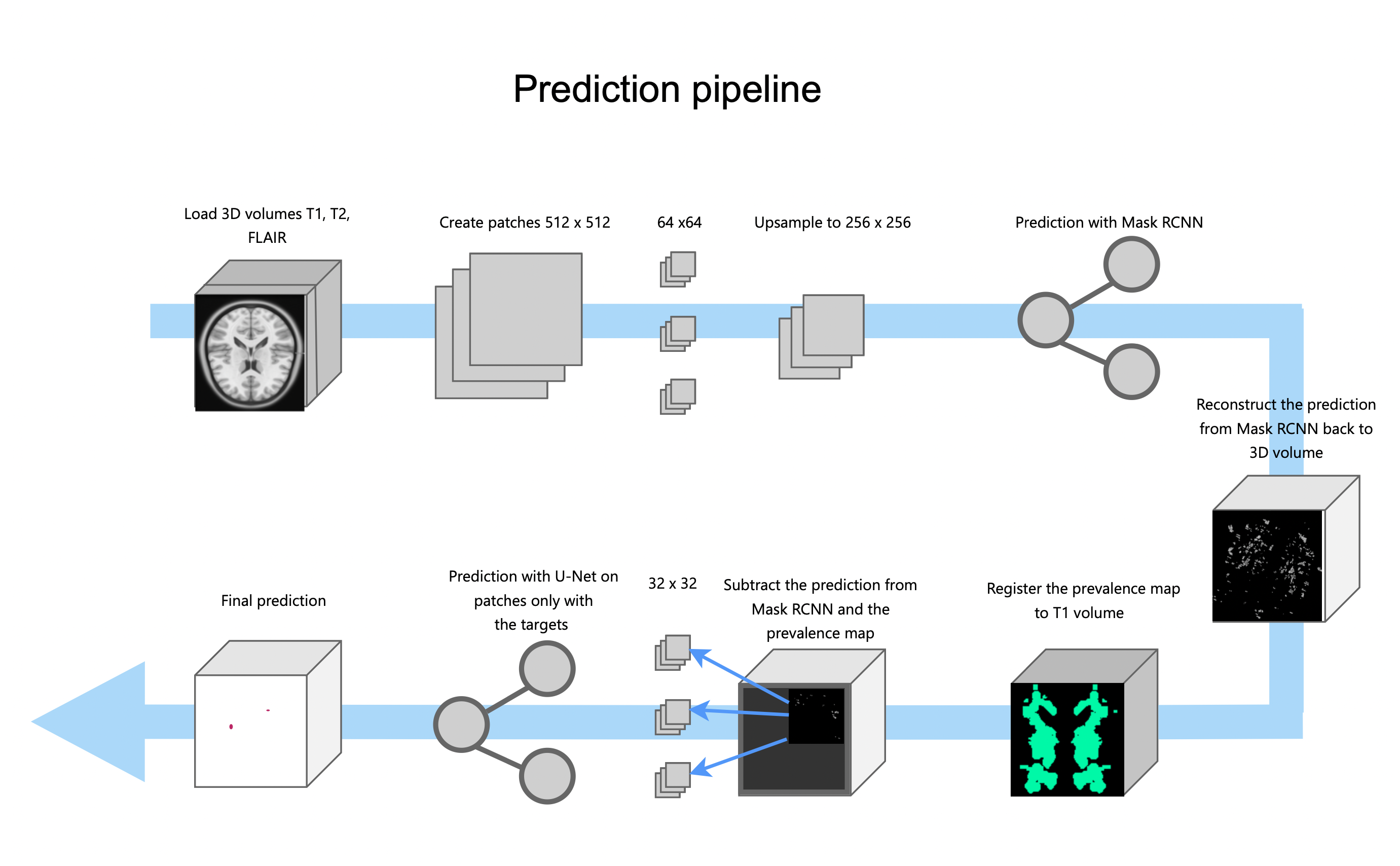}
\caption{The full prediction pipeline showing the pre-processing steps, the two stages (Mask R-CNN and U-Net) and the application of the lesion prevalence map; to generate the final lacune segmentation for a patient.}
\label{fig:fig2}
\end{figure}

The full prediction pipeline is shown in Figure \ref{fig:fig2}. The various steps of the pipeline are: z-score intensity normalization, selecting slices and patches for Mask R-CNN, up-sampling the 64$\times$64 patches to 256$\times$256, applying Mask R-CNN, down-sampling back to 64$\times$64 and reconstructing the full 3D image, applying the lacune prevalence map to remove false positive detections, creating 32$\times$32 patches for the remaining detections, segmenting them with the U-Net, and finally reconstructing the final segmentation map. The following sections describe this pipeline in detail.

\subsubsection{Pre-processing}
All image data was scaled using z-score normalization (zero-mean-unit-variance) on the patient level.

\subsubsection{Stage 1: Mask R-CNN}
Mask R-CNN was trained on 2D patches of size 64$\times$64 with 50\% overlap, extracted from all slices with lacunes in the dataset. A pretrained Mask R-CNN with a Resnet50 backbone trained on ImageNet was selected. This network is more familiar with objects of medium and large size; and empirical results showed that the current implementation cannot detect the lacunes, because of their small size. In order to detect lacunes with this network, patches were up-sampled with nearest neighbour interpolation by a factor four to a size of 256$\times$256. This creates object sizes in which the target is larger and more similar to the pre-trained network.

Additional parameters of the Mask R-CNN include the anchor size and aspect ratios. The chosen anchor size ranged from 4 to 64 and the aspect ratios were 0.02, 0.25, 1.0, 2.0, and 2.75. The T1, T2, and FLAIR images were provided as three separate channels to the network with a batch size of six. Data augmentation consisted of random horizontal flips. The model was trained for 20 epochs, after which there was no apparent improvement.

\subsubsection{Stage 2: U-Net}
A U-Net was employed to segment lacunes at every location detected by Mask R-CNN in stage one. The U-Net was trained on 2D patches of 32$\times$32 with 50\% overlap, selected from the locations within the lesion prevalence map in 10/90 ratio for lacune/background (selected empirically on validation data). The model was trained for 30 epochs, and visual inspection showed that after 25 epochs there was no improvement on the training and validation losses (see Results Figure \ref{fig:fig3}).

\subsubsection{Applying the lesion prevalence map}
The lesion prevalence map was registered to the T1-weighted images using elastix \cite{Klein2010}, using an affine transformation followed by a bspline transformation. The lesion prevalence map was used to mask out false positive segmentations.   

\subsubsection{Uncertainty}
The VALDO challenge requires an additional uncertainty map to be submitted. Limited time and resources prohibited us from computing the actual epistemic or aleatoric uncertainty. Prior experience suggests that methods are most uncertain at the boundaries of objects. Visual inspection of our results suggest that the final method tends to under-segment lacunes. Therefore we decided to create a 1-pixel in-plane border outside of the segmented lacunes, as a pseudo-uncertainty.

\subsection{Evaluation }
\subsubsection{Stage 1: Mask R-CNN}
The goal of the Mask R-CNN was to localize all the lacunes (high sensitivity) with a trade-off of potential low specificity. The model produced two outputs that were taken into account, the bounding-box and segmentation predictions; which were both optimized. To evaluate these predictions, we calculated three measures between the bounding-box predictions and the bounding-box of the ground truth; and also between the segmentation predictions and the segmentation ground truths. The measures were as follows: Intersection over Union (IoU), Average Precision, and Average Recall. Performance of the Mask R-CNN is defined with these three measures, calculated separately for areas of small, medium, and large size, and all combined. In other words, the model gives a clear evaluation for different areas based on their size. For example, the model with both high Average Precision and Average Recall on small areas, can accurately localize small lacunes.

\subsubsection{Stage 2: U-Net}
The goal of the U-Net was to segment lacunes in the ROIs given by the Mask R-CNN. To evaluate the U-Net, we calculated the DICE between the segmentations and the ground truth. We optimized the threshold for the posterior probability based on the DICE of the validation set.

\section{Results}
\subsection{Stage 1: Mask R-CNN}
The best results of the Mask R-CNN were achieved after 20 epochs, both for the bounding box and segmentation predictions.  The highest value for the Average Precision and Average Recall was measured on areas of medium size; both for the bounding box and segmentation predictions. The Average Precision for the medium-size bounding box was 0.02 and for segmentation was 0.06. The Average Recall for the  medium-size bounding box was 0.23 and for segmentation 0.27. This suggests that the model might be best in localizing medium-sized lacunes, than small or large lacunes. Visual inspection showed that the majority of lacunes were correctly localized, although at the cost of a large number of FPs.

\subsection{Stage 2: U-Net}
\begin{figure}[tbh]
\centering
\includegraphics[width=\textwidth]{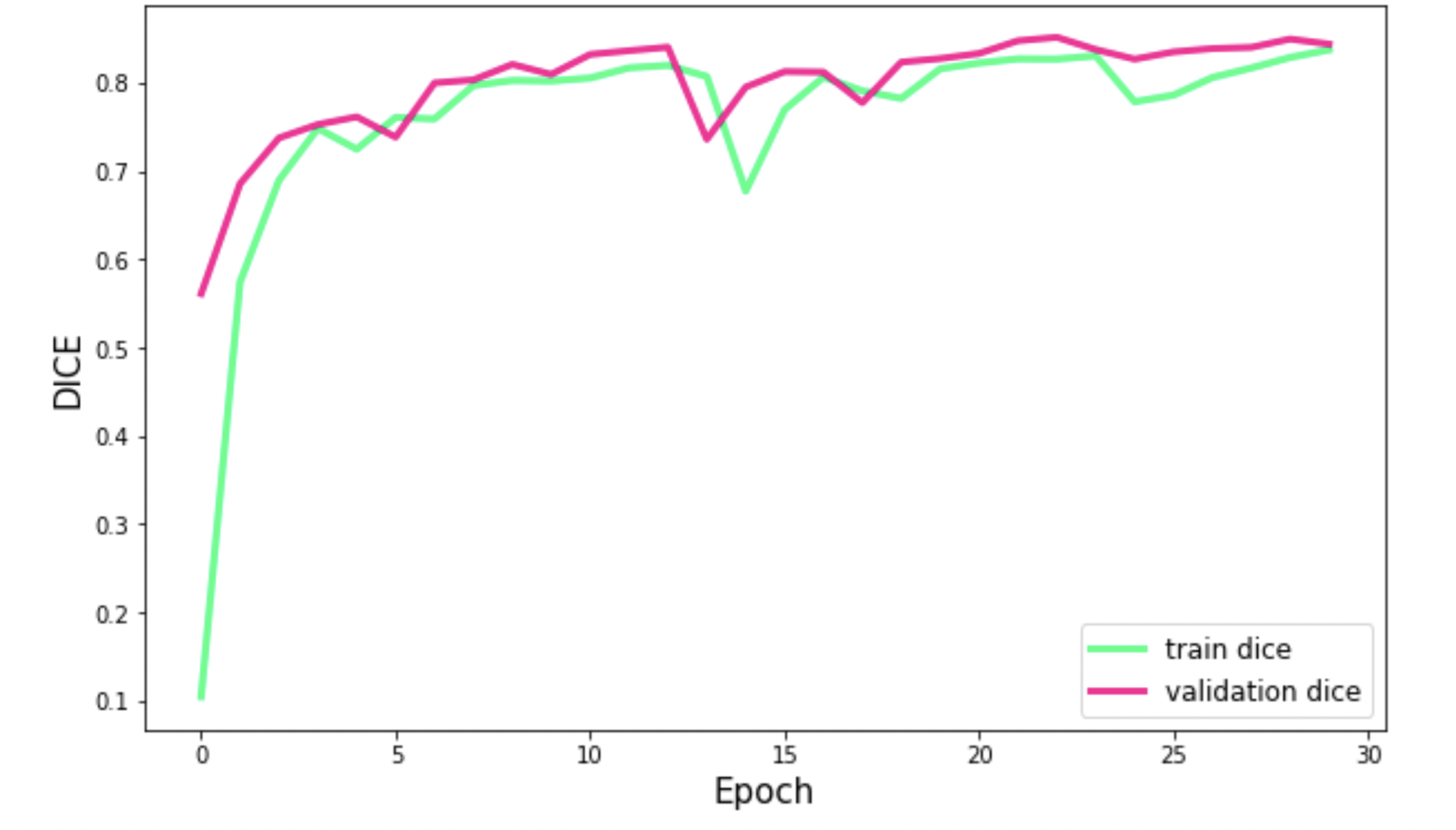}
\caption{Progression of the Dice metric during training, reported on both the training and validation set.}
\label{fig:fig3}
\end{figure}

The U-Net model was trained and evaluated independently from the Mask R-CNN in three different dataset splits. The first case was on a 50/50 split for lacune/background, resulting in a mean DICE on the validation set of 0.87. However, this model predicted many false positives (FPs). For example, the model predicted parts with low-intensity values, like the CSF, to be a lacune. For this reason, more slices without lacunes were included in the training process, with a new split of 25/75 for lacune/background. The given ratio improved the results by having a visual inspection. Despite that still, many FPs remained. Lastly, the dataset with a split of 10/90 for the lacune/background can recognize the lacunes and produced only a minimal number of FPs, which were removed with the prevalence map. We report the mean DICE on the training set of 0.83 and on the validation set of 0.84. The training loss got down to 0.19145 after 30 epochs and the evaluation metrics during training can be seen in Figure \ref{fig:fig3}.

\subsection{Qualitative evaluation}
\begin{figure}[tbh]
\centering
\includegraphics[width=\textwidth]{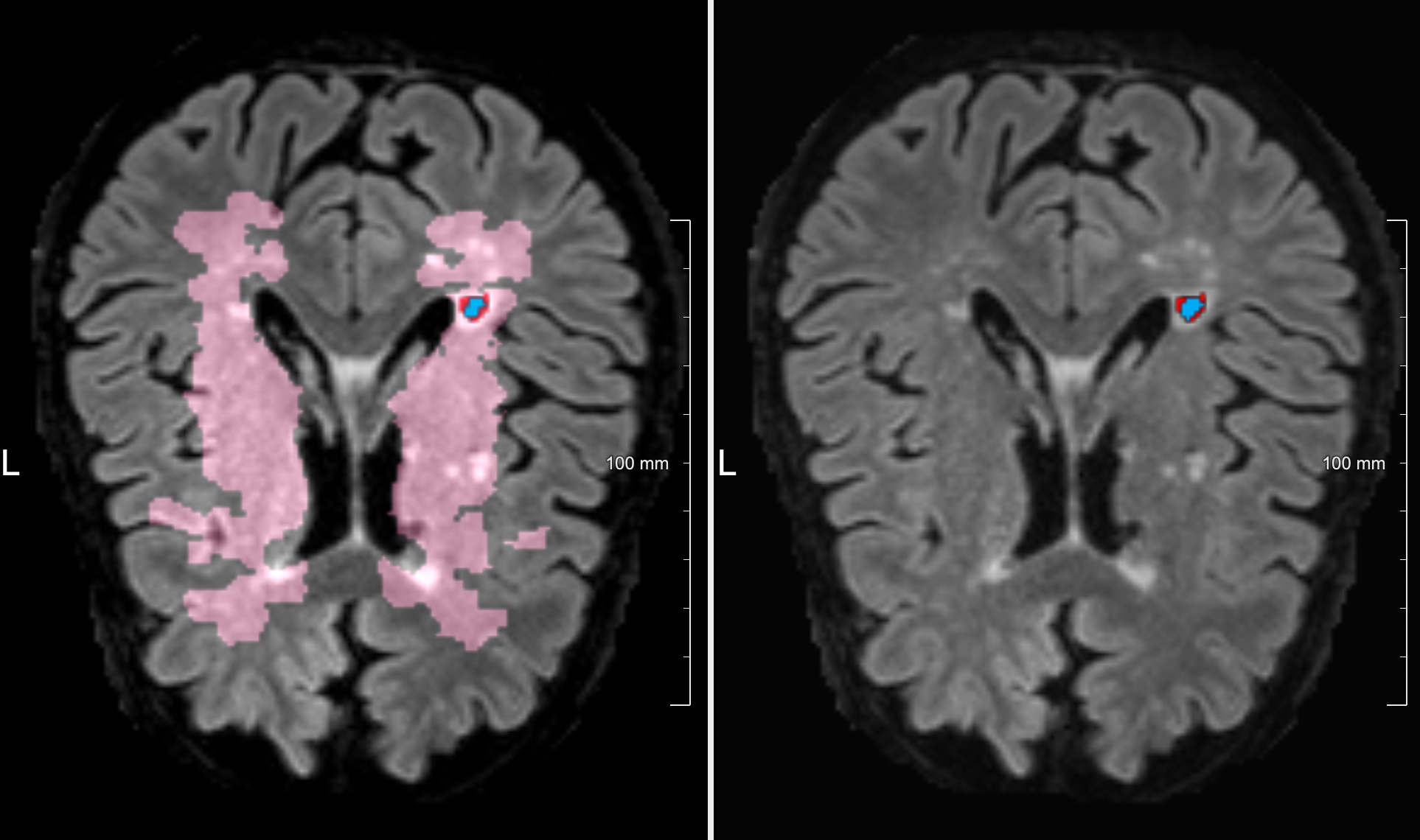}
\caption{Example of a correctly detected but slightly under-segmented lacune. Pink is the lacune prevalence map, red is the ground truth segmentation, and blue is the predicted segmentation.}
\label{fig:fig5}
\end{figure}

Visual inspection of the results showed that most of the lacunes were correctly detected. An example is shown in Figure \ref{fig:fig5}. However, the predicted binary mask does not precisely match with the ground truth. In some cases, the prediction was a bit smaller compared to its ground truth. A small number of false positive detections remain.

\begin{figure}[p]
\centering
\includegraphics[width=\textwidth]{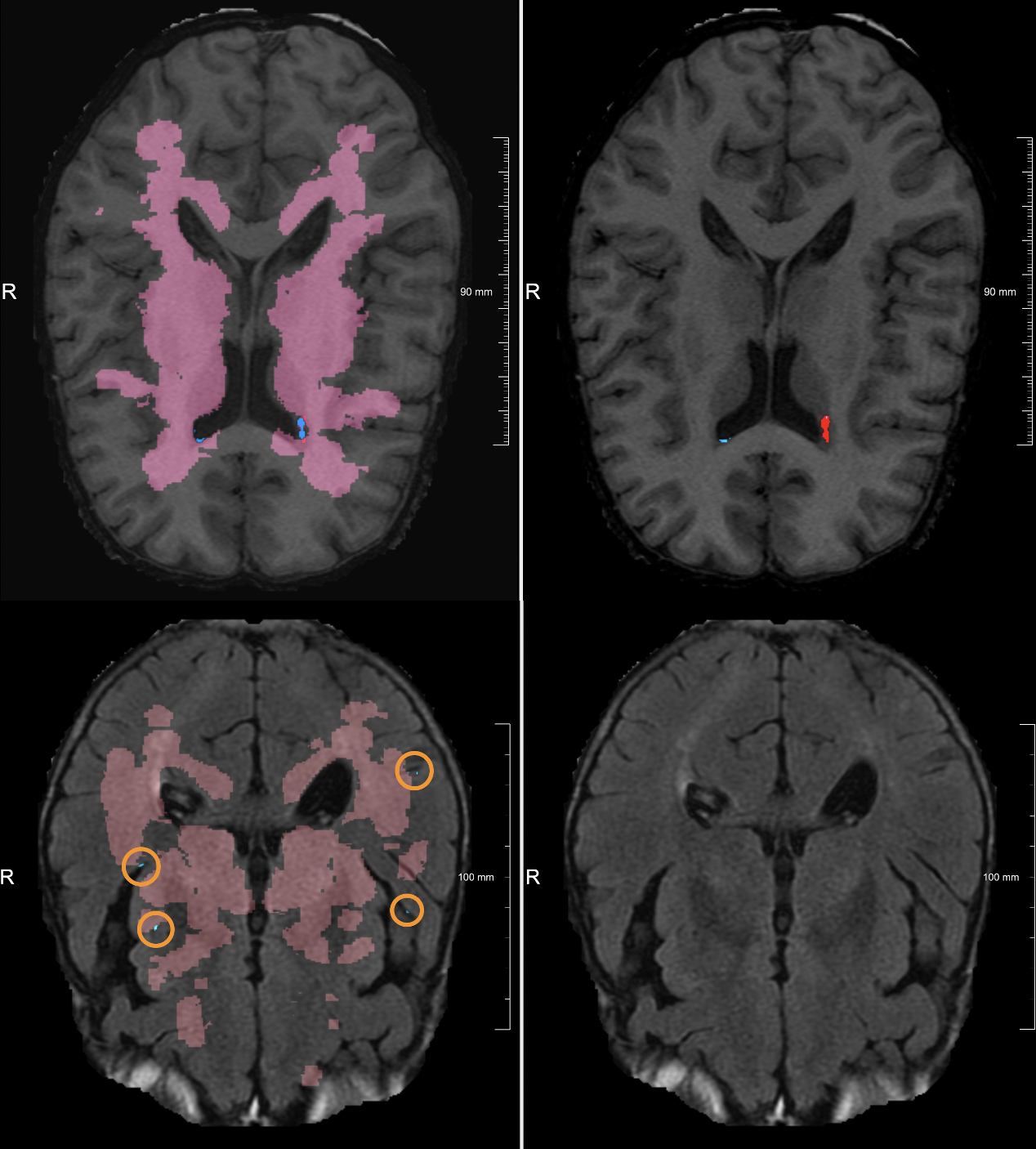}
\caption{The first pair of images (top row) shows a correctly segmented lacune which was later removed by the prevalence map. The second pair (bottom row) shows how false positives are removed by the prevalence map.}
\label{fig:fig4}
\end{figure}

Figure \ref{fig:fig4} shows the positive and negative effects of the lacune prevalence mask. Most of the times, the lesion prevalence map correctly removes false positive detections; as can be seen in the bottom row of Figure \ref{fig:fig4}. Unfortunately, sometimes the lesion prevalence map also removes a true positive; as can be seen in the top row of Figure \ref{fig:fig4}. This is most likely caused by an image registration error, where the close proximity of the lacune to the ventricles caused a misalignment.

\section{Discussion}
In this work, we presented a two-stage approach to segment lacunes of presumed vascular origin on brain MR images. The results suggest that this method can accurately detect most of the lacunes on our training/validation dataset. As illustrated in Figure \ref{fig:fig2}, the U-Net decides whether the input from the Mask R-CNN is the true lacune or not and produces the final binary segmentation. 

The Mask R-CNN currently produces a substantial number of false positive detections. On the one hand, a high sensitivity is essential in this first stage, probably at the cost of a moderate specificity. On the other hand, a reduction of the number of false positives in this first stage might help the overall prediction pipeline. Future work could look into thresholding the Mask R-CNN results, to provide a better selection of lacune candidates to stage two.

A limitation of our U-Net model is the use of 2D patches. In some cases, the lacune was correctly segmented on one slice, but missed on the next slice. Using a 3D U-Net might solve this issue, because it can consider the full 3D extent of the lacunes.

The use of a lacune prevalence map seems to improve the method results, by removing a substantial amount of false positive detections; at the cost of also removing some true positives. For future applications, the use of a lacune prevalence map has some limitations. Even though lacunes occur at preferential locations, they can occur throughout the full brain. Using a map to mask out detections, might introduce false negatives at rare locations.

In conclusion, a two-stage deep learning method shows promise for facilitating detection and segmentation lacunes of presumed vascular origin.

\section{More information}
Source code is available at: \url{https://github.com/hjkuijf/MixLacune}. The docker container hjkuijf/mixlacune can be pulled from \url{https://hub.docker.com/r/hjkuijf/mixlacune}.

\bibliographystyle{ieeetr} 
\bibliography{}

\end{document}